\documentclass{interact}

\usepackage{graphicx}
\usepackage{epstopdf}
\usepackage{braket}

\usepackage[numbers,sort&compress,merge]{natbib}
\bibpunct[, ]{[}{]}{,}{n}{,}{,}

\newcommand{\rmd}{\mathrm{d}}

\newcommand{\etal}{\emph{et al.}}

\begin{document}

\articletype{SPECIAL ISSUE IN HONOR OF ANDR\'{E} D. BANDRAUK}

\title{Quantum stability of an ion in a Paul trap revisited}

\author{\name{A. Hashemloo and C. M. Dion\thanks{CONTACT C.~M. Dion Email: claude.dion@umu.se}}
  \affil{Department of Physics, Ume{\aa} University, SE-901\,87 Ume{\aa},
    Sweden}}

\maketitle

\begin{abstract}
  We study the quantum stability of the dynamics of ions in a
  Paul trap.  We revisit the results of Wang \etal\ [Phys. Rev. A
  \textbf{52}, 1419 (1995)], which showed that quantum trajectories
  did not have the same region of stability as their classical
  counterpart, contrary to what is obtained from a Floquet analysis of
  the motion in the periodic trapping field.  Using numerical
  simulations of the full wave-packet dynamics, we confirm that the
  classical trapping criterion are fully applicable to quantum motion,
  when considering both the expectation value of the position of the
  wave packet and its width.
\end{abstract}



\section{Introduction}

Paul traps are devices using a radio-frequency, time-de\-pen\-dent
electric field to confine charged
particles~\cite{Paul_RMP_1990,Ghosh_book_1995,Major_book_2005}.  With
the right combination of electric potential strength and field
frequency, one or many trapped particles will move in closed
trajectories inside the trap.  Such devices have been used for
instance to trap atomic and molecular ions, for studies ranging from mass spectrometry~\cite{March_MSR_2009}, high-precision spectroscopy~\cite{Leanhardt_JMS_2011} and ultracold chemistry~\cite{Drewsen_PRA_2000,Baba_JCP_2002,Blythe_PRL_2005,Tong:CPL_2012,Rosch_JCP_2014} to quantum state manipulation~\cite{Leibfried_RMP_2003,Leibfried_NJP_2012} and quantum simulations~\cite{Schneider_RPP_2012}, and recently tests of the time variation of fundamental constants~\cite{Huntemann_PRL_2014}.

Of importance is the question of stability.  As mentioned above, the
motion of a charged particle must be constrained for the particle to
be actually trapped.  This will depend in particular on the relative
electric potentials applied to the electrodes making up the trap and
the frequency of the oscillating field.  For a classical particle, the
motion is given by a Mathieu equation~\cite{Wolf_NIST_2010}, for which
stable solutions correspond to cases where the trajectory is bounded.
For a quantum particle, the situation is more complicated, but a
Floquet
analysis~\cite{Combescure_AIHPA_1986,Brown_PRL_1991,Glauber_1992,Glauber_1992b}
shows that the average motion of the charged particle follows exactly
the classical Mathieu equation, implying that the classical stability
criterion also applies in the quantum case.  Such a conclusion is also
supported by other theoretical
derivations~\cite{Li_PRA_1993,Nieto_NJP_2000,Hai_JPB_2003} and
numerical simulations of the wave-packet
dynamics~\cite{Hashemloo_IJMPC_2016}.  While these studies pertain to
atomic ions, non-polar molecules have the same centre-of-mass
trajectories as atomic ions, and we have also recently shown that the
interaction between the trapping field and a dipole moment doesn't
lead to a noticeable modification of the stability
criterion~\cite{Hashemloo_JCP_2015}. 

However, a paper by Wang \etal~\cite{Wang_PRA_1995} purports to show
that there are trapping parameters for which a quantum trajectory is
stable, while the classical counterpart would not be.  Their analysis
was based on a study of the convergence of a function series of a
dynamical variable related to the wave function of the charged
particle.  We revisit here this claim by performing a direct
simulation of the wave-packet dynamics for the same trapping
conditions as considered by  Wang \etal, as well as by comparing with
the corresponding classical trajectories.

This paper is arranged as follows.  We start by presenting the
trapping model, based on Reference~\cite{Wang_PRA_1995}, and the
corresponding classical stability criterion. The numerical methods
corresponding to both quantum and classical simulations are presented
in Section~\ref{sec:num}.  This is followed, Section~\ref{sec:results}, by
the results of the various simulations, for stable and unstable
trapping parameters.  Finally, concluding remarks are given in
Section~\ref{sec:conclusion}.

\section{\label{sec:model}Model for an ion in a Paul trap}

\subsection{Trapping potential}

We consider an ion of charge $e$ and mass $m$ in a two-dimensional
Paul trap with the trapping potential~\cite{Wang_PRA_1995}
\begin{equation}
  V(x,z) = \frac{e}{2 r_0^2} \left( U_0 + V_0 \cos \omega t \right)
  \left(x^2 - z^2 \right),
\label{eq:trap_wang}
\end{equation}
where $r_0$ is the radius of the trap, $U_0$ and $V_0$ the electric
potential on the static and radio-frequency electrodes, respectively,
with $\omega$ the frequency of the time-dependent electric potential.
From this point on, we take units such that $e = \hbar = m = r_0 =1$,
and will consider the case $\omega = 5$, the same as for the main
results of Reference~\cite{Wang_PRA_1995}.

\subsection{Classical trajectories and stability}
\label{sec:stability}

The classical equations of motion for the ion are given here by
\begin{eqnarray} 
  \ddot{x} + \left( U_{0} + V_0 \cos \omega t  \right) x &=& 0, \nonumber \\
  \ddot{z} - \left( U_{0} + V_0 \cos \omega t  \right) z &=& 0.
  \label{eq:eq_motion_2D}
\end{eqnarray}
Using the substitutions
\begin{equation}
    a_{x,z} \equiv \pm \frac{4}{\omega^2} U_0, \quad
    q_{x,z} \equiv \mp \frac{2}{\omega^2} V_0, \quad
    \tau \equiv \Omega t/2,
    \label{eq:wang_aq}
  \end{equation}
allows us the rewrite Equations~(\ref{eq:eq_motion_2D}) as to 
Mathieu equations for $\mathbf{r} \equiv (x,z)$,
\begin{equation} 
  \frac{\rmd^{2} \mathbf{r} }{\rmd \tau^{2} } + (a - 2q\cos
  2\tau )\mathbf{r} = 0. 
  \label{eq:Math.eq}
\end{equation}
Bounded solutions, where $\mathbf{r}$ remains finite, of the Mathieu
equation exist for certain regions in the $a$--$q$
plane~\cite{Wolf_NIST_2010}.  Correspondingly, for certain values of
the electric potentials $U_0$ and $V_0$, the ion will remain inside
the trap (neglecting possible collisions with the trap's electrodes if
the amplitude of the motion is too big). This condition is commonly
considered the \emph{classical stability region} of an ion
trap~\cite{Paul_RMP_1990,Ghosh_book_1995,Major_book_2005}, and is the
criterion we are using in this paper to label trapping potentials as
stable.  The boundaries in the $a$--$q$ plane between stable and
unstable regions are given by the Mathieu characteristic values $a_r$
and $b_r$~\cite{Wolf_NIST_2010}, and the resulting stability diagram
is plotted in Figure~\ref{fig:stability_wang}.
In this region of the $(V_0,U_0)$ plane, the stability border in $x$
is obtained from $a_1$, while the one in $z$ from $a_0$.  The crossing
point is found at $U_0 = 1.48121$, $V_0 = 8.82495$.

\section{Numerical methods}
\label{sec:num}

We present here a brief description of the simulations of the dynamics
of the trapped ion, for both the quantum and classical cases.  Further
details of the numerical approaches used can be found in
Reference~\cite{Hashemloo_IJMPC_2016}.

\subsection{Quantum mechanical approach}
\label{sec:num_quant}

For the quantum wave packet dynamics, we have used the program
\textsc{wavepacket}~\cite{Dion_CPC_2014}, which is based on the
split-operator
method~\cite{Feit_JComputP_1982,Feit_JCP_1983,Bandrauk_CPL_1991}, with
the wave function expressed on a discrete set of grid points.  The
Hamiltonian for the motion of the ion is simply
\begin{equation}  
  \hat{H} = -\frac{1}{2} \nabla_{\mathbf{r}}^{2} + V(\mathbf{r}),
  \label{eq:hamiltoni}
\end{equation}
with $V(\mathbf{r})$ the potential given by Equation~(\ref{eq:trap_wang}). While the system considered is two-dimensional, the equations of motion in $x$ and $z$ are separable, and the total wave function is obtained from the combination of two one-dimensional simulations.

To compare the quantum dynamics to the corresponding classical
trajectories, we take the initial wave function to be a Gaussian,
\begin{equation}
  \psi(x,z; t=0) = \frac{1}{\sqrt{\pi \sigma_x \sigma_z}}
  \exp \left[ -\frac{ \left( x - x_0 \right)^2}{2
      \sigma_x^2} \right] \exp \left[ -\frac{ \left( z - z_0
      \right)^2}{2 \sigma_z^2} \right]
  \label{eq:gaussian2D}
\end{equation}
with $x_0=z_0=1$.  We choose the widths $\sigma_x = \sigma_z = 2$, for
which the spreading of the wave packet will be
reasonable~\cite{Hashemloo_IJMPC_2016}.  We use 393216 grid points in
the range $[-600,600]$ along both the x and z axes for the simulations
within the stability region, and increase the grid to $[-1600,1600]$
with 1600000 points for the unstable trajectories.  The time step used
is $\Delta t = 5 \times 10^{-4}$.

\subsection{Classical approach}
\label{sec:num_class}

The expectation values of position $\braket{x}$ and $\braket{z}$ of
the wave packet of the trapped ion can be compared with its classical
analogue, given by the Mathieu equations~(\ref{eq:Math.eq}).  We
obtain the classical dynamics by a direct integration of the classical
equations of motion~(\ref{eq:eq_motion_2D}), using the symplectic
St\"{o}rmer-Verlet integrator~\cite{Hairer_book_2006}.

To second order in the time step $\Delta t$, the phase-space dynamics
for the momentum $\mathbf{p}_n$ and position $\mathbf{r}_n$ at the
$n^{\mathrm{th}}$ time step are obtained
from~\cite{Hashemloo_IJMPC_2016,Hairer_book_2006}
\begin{eqnarray} 
  \mathbf{p}_{n + \frac{1}{2}} &=& \mathbf{p}_{n} - \frac{\Delta t}{2}
    \mathbf{H}_{\mathbf{r}_n}, \nonumber \\ 
    \mathbf{r}_{n + 1} &=& \mathbf{r}_{n} + \Delta t
    \frac{\mathbf{p}_{n + \frac{1}{2}}}{m}, \nonumber \\ 
    \mathbf{p}_{n + 1} &=& \mathbf{p}_{n + \frac{1}{2}} - \frac{\Delta
      t}{2} \mathbf{H}_{\mathbf{r}_{n+1}}, 
  \label{eq:StormerVerlet}
\end{eqnarray}
where $\mathbf{H}_{\mathbf{r}_{n}}$ is the vector of partial
derivatives of the Hamiltonian $H(\mathbf{p},\mathbf{q})$ with respect
to the components of the position $\mathbf{r}$,
\begin{eqnarray}
  H_{x_n} &=& \left[ U_0 + V_0 \cos ( \omega t_n) \right] x_n,
  \nonumber \\
  H_{z_n} &=& -\left[ U_0 + V_0 \cos ( \omega t_n) \right] z_n.
\end{eqnarray}
The system of equations~(\ref{eq:StormerVerlet}) is iterated starting
from an initial condition equivalent to the quantum simulation (see
Section~\ref{sec:num_quant}), namely $\mathbf{r}_0 = (1,1)$ and
$\mathbf{p}_0 = 0$, with a time step $\Delta t = 5 \times 10^{-4}$.

\section{Results}
\label{sec:results}

\begin{figure}
  \begin{center}
    \includegraphics[width=0.65\textwidth]{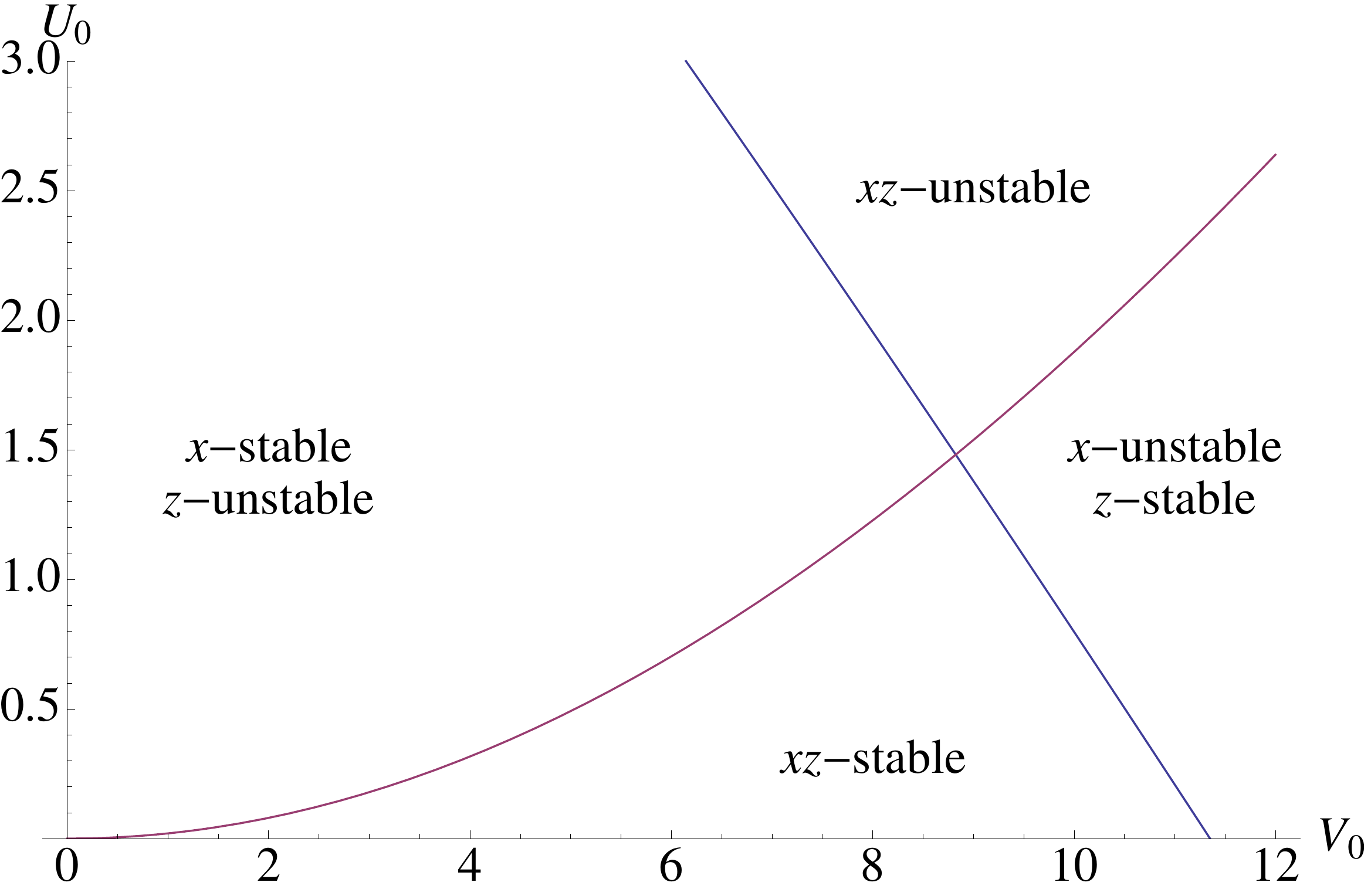} 
  \end{center}
\caption{\label{fig:stability_wang}Stability diagram
  for the Mathieu equation~(\ref{eq:Math.eq}) with parameters $a$ and
  $q$ from Equations~(\ref{eq:wang_aq}), calculated for $\omega = 5$ in
  units where $e = \hbar = m = r_0 = 1$.}
\end{figure}

Comparing with Figures~1 and 2 of Reference~\cite{Wang_PRA_1995}, we see that
the entire region labelled $xz$-unstable in
Figure~\ref{fig:stability_wang} is, according to Wang \etal, stable for
quantum dynamics, since their stability region extends to $U_0 \sim 
3.3$. We therefore compare two trapping conditions, one below the
crossing between the classically stable and unstable regions (see
Section~\ref{sec:stability}), the other above.

\begin{figure}
  \begin{center}
    \includegraphics[width=0.99\textwidth]{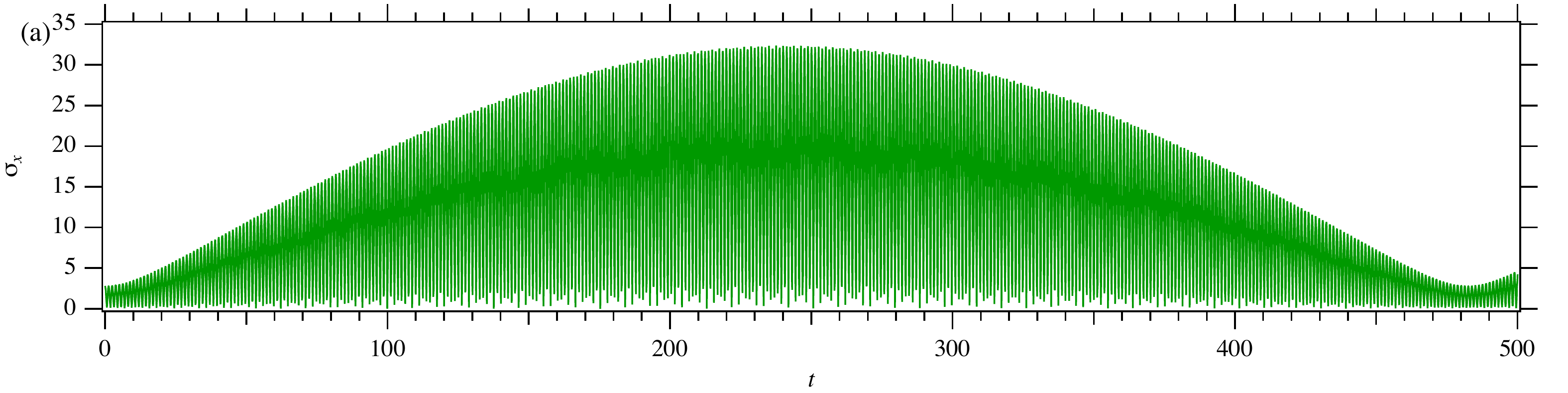}
    \includegraphics[width=\textwidth]{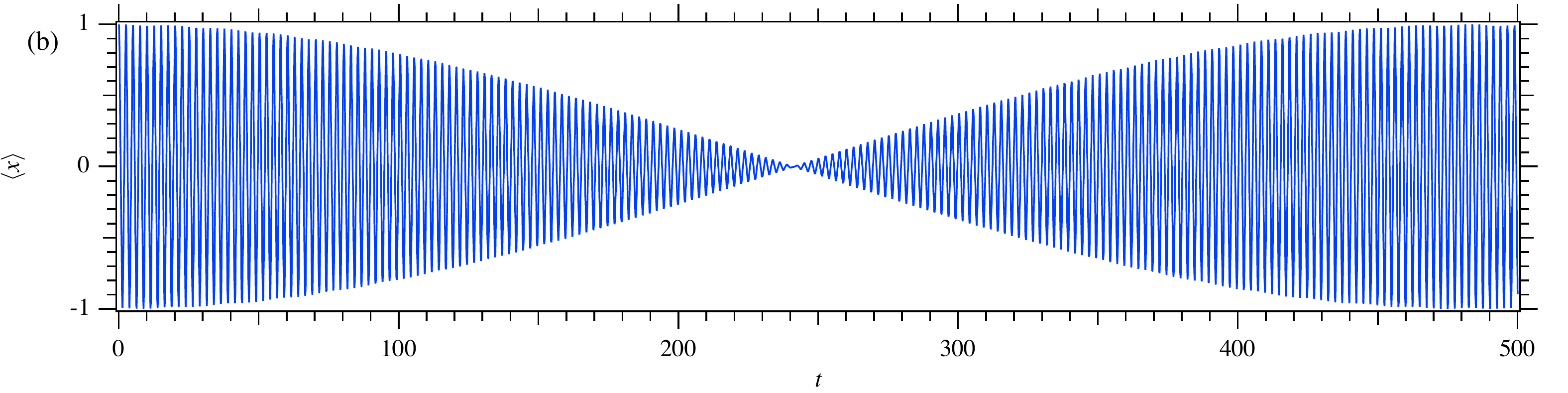}
    \includegraphics[width=\textwidth]{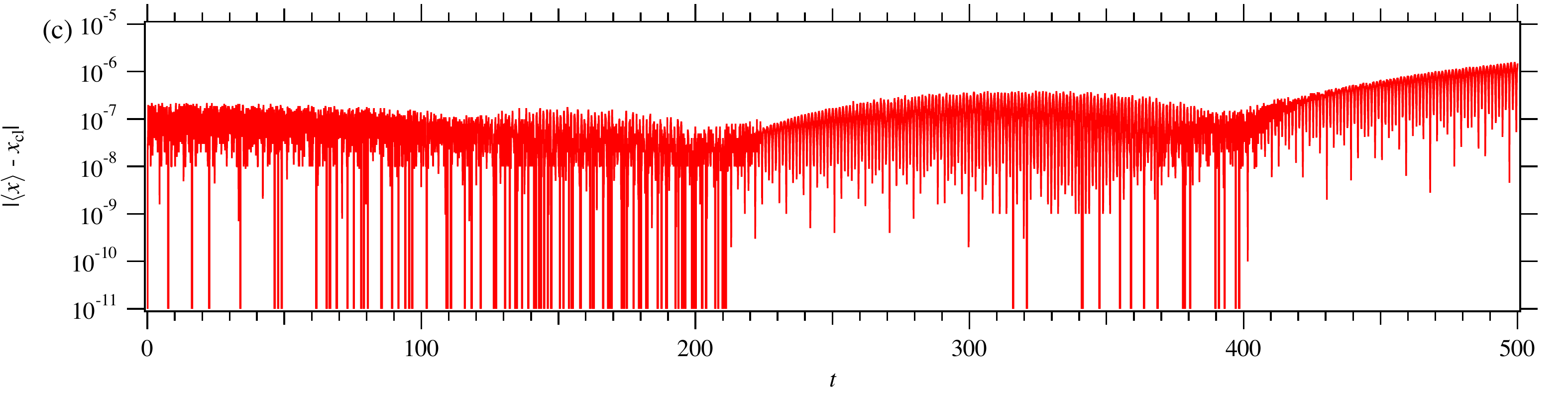}
  \end{center}
  \caption{\label{fig:stable_x}Time evolution of the
    ion's wave packet along the $x$ axis, for trap parameters $U_0 =
    1.4811$ and $V_0 = 8.825$, and $\omega = 5$.  (a) Wave-packet
    width $\sigma_x$; (b) expectation value of the position
    $\braket{x}$; (c) difference between quantum and classical
    trajectories.}
\end{figure}

\begin{figure}
  \begin{center}
    \includegraphics[width=0.99\textwidth]{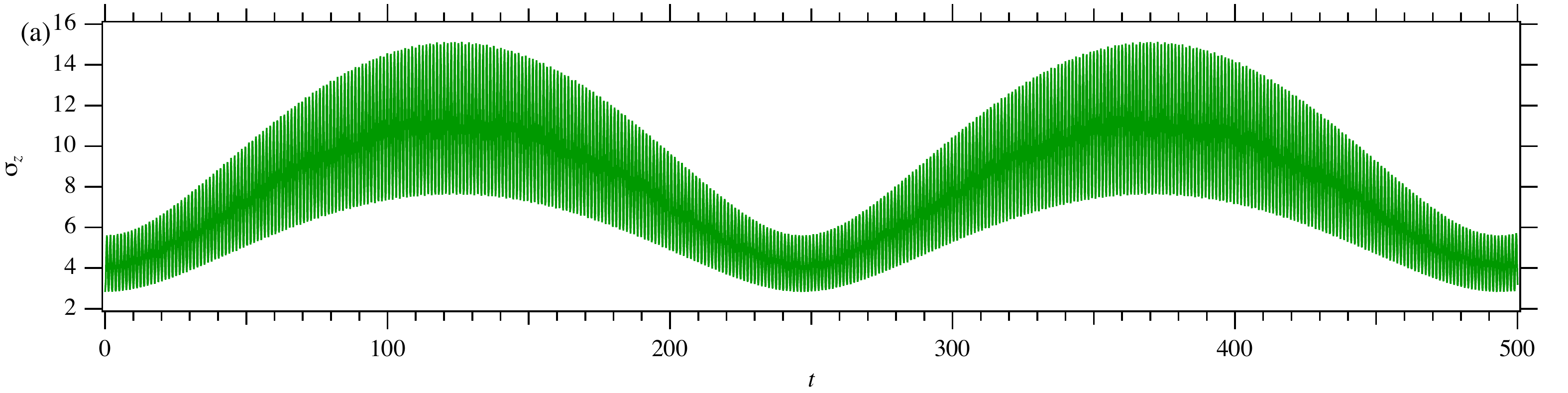}
    \includegraphics[width=\textwidth]{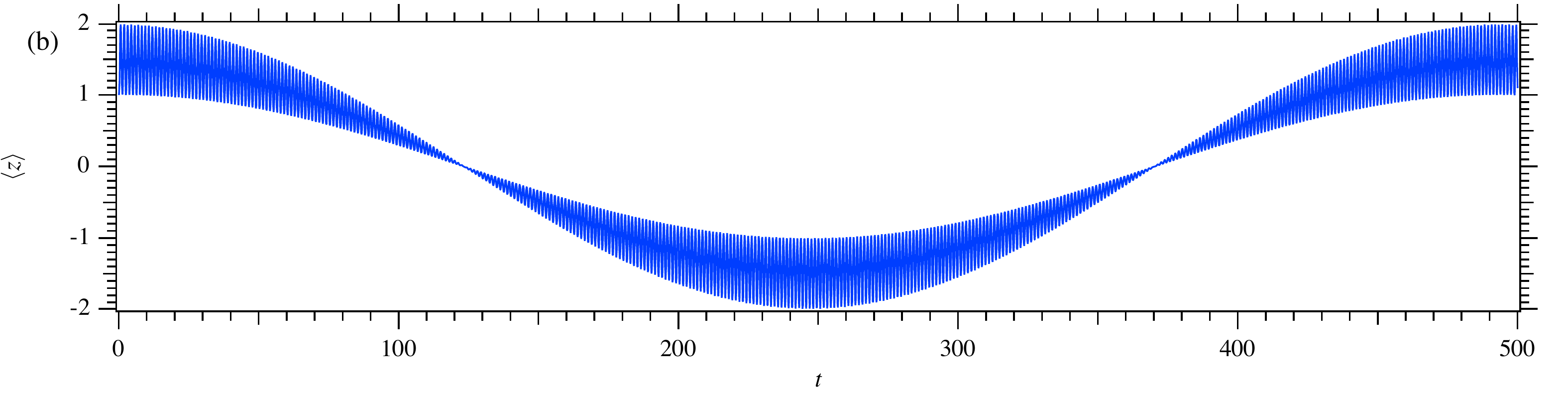}
    \includegraphics[width=\textwidth]{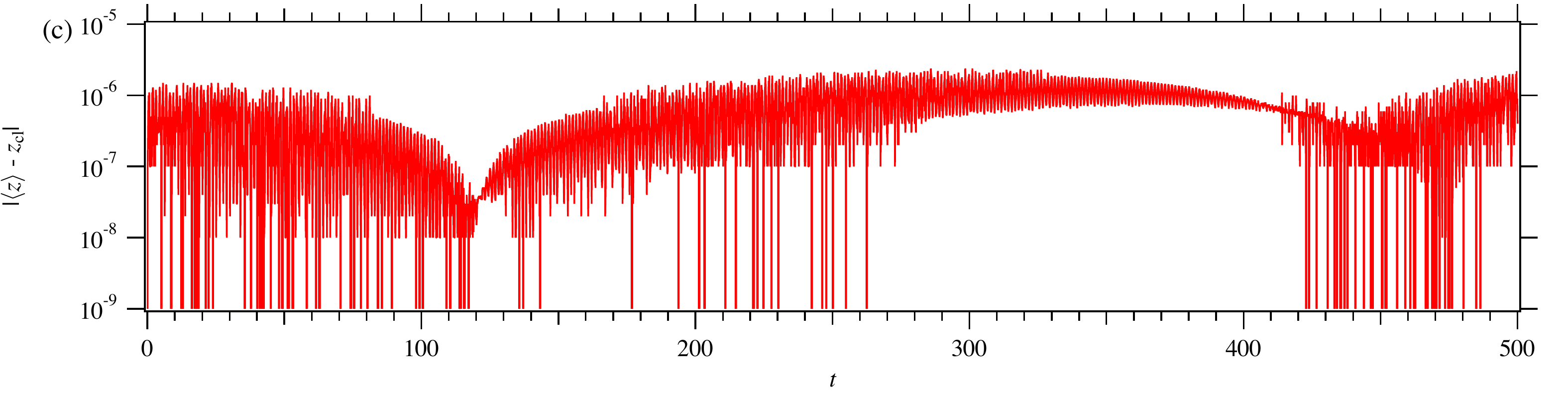}
  \end{center}
  \caption{\label{fig:stable_z}Time evolution of the
    ion's wave packet along the $z$ axis, for trap parameters $U_0 =
    1.4811$ and $V_0 = 8.825$, and $\omega = 5$.  (a) Wave-packet
    width $\sigma_z$; (b) expectation value of the position
    $\braket{z}$; (c) difference between quantum and classical
    trajectories.}
\end{figure}

Taking parameters $\omega = 5$, $U_0 = 1.4811$ and $V_0 = 8.825$,
which corresponds to stable trajectories according both to the
classical analysis (Figure~\ref{fig:stability_wang}) and that of
Reference~\cite{Wang_PRA_1995}, we find indeed a periodically oscillating,
bound trajectory and width of the quantum wave packet, see
Figures~\ref{fig:stable_x} and \ref{fig:stable_z}.
There is also a perfect accord (within numerical error) between the
classical trajectory and the expectation value of position for the
quantum simulation, as calculated by the absolute difference 
\begin{equation}
  \left| \braket{x} - x_\mathrm{cl} \right|
\end{equation}
where $x_\mathrm{cl}$ is the position in the classical trajectory (see
Section~\ref{sec:num_class}), with an equivalent equation in $z$, as
shown in Figures~\ref{fig:stable_x}(c) and \ref{fig:stable_z}(c).

\begin{figure}
  \begin{center}
    \includegraphics[width=0.99\textwidth]{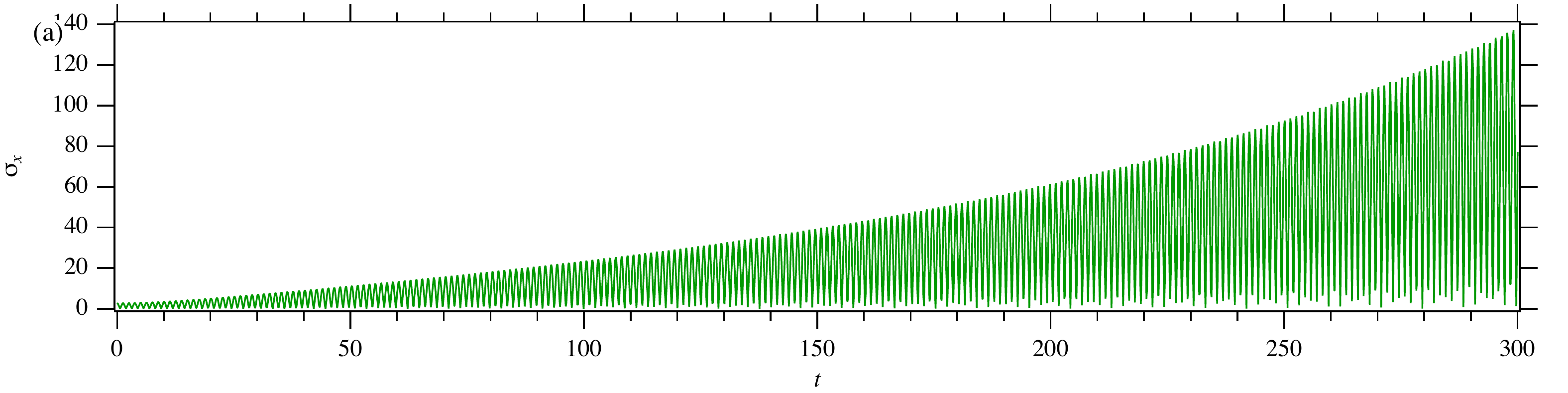}
    \includegraphics[width=\textwidth]{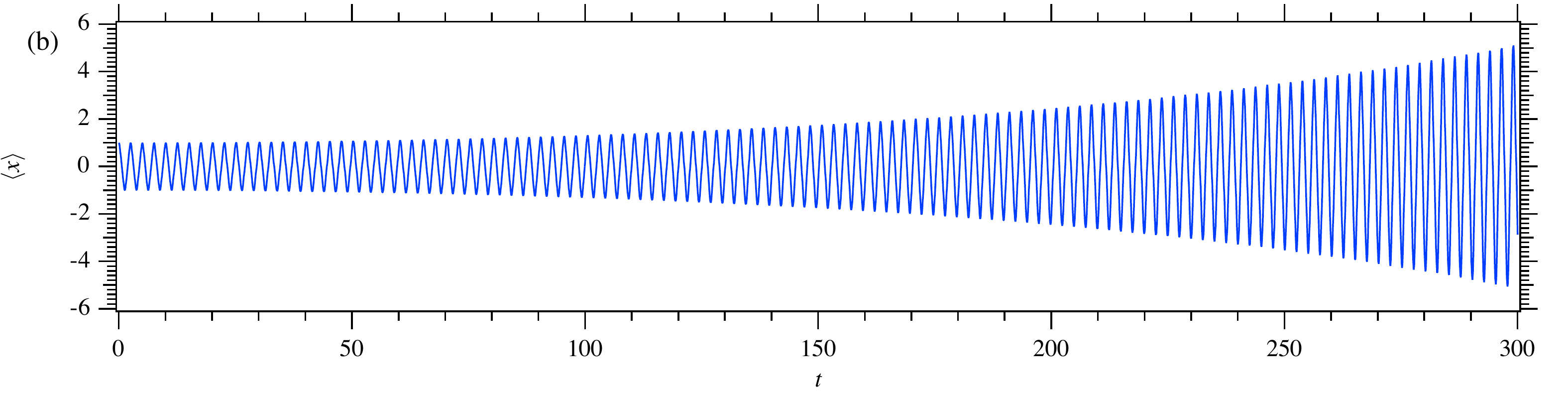}
    \includegraphics[width=\textwidth]{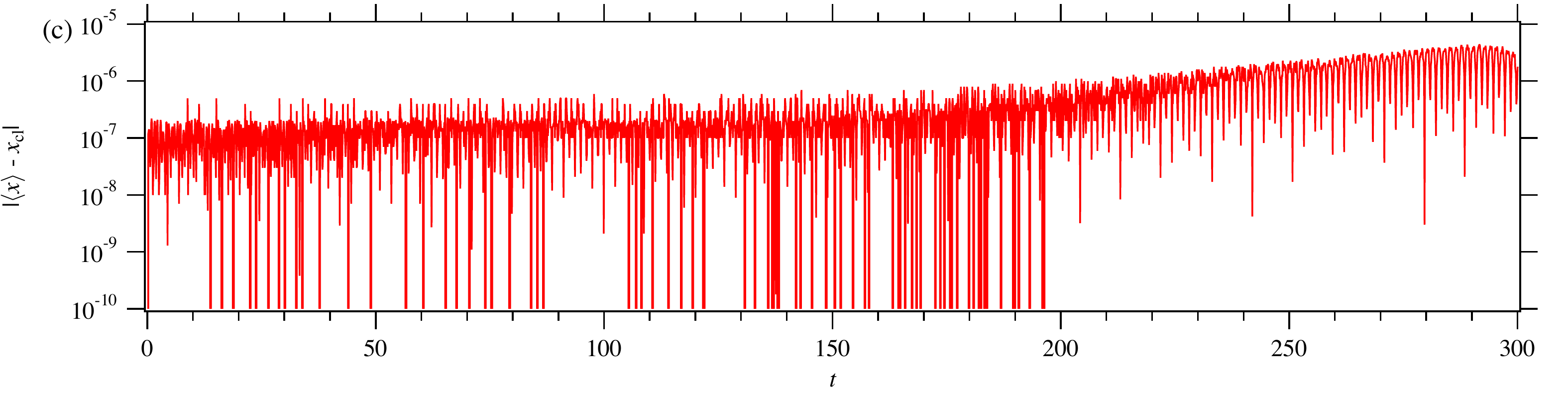}
  \end{center}
  \caption{\label{fig:unstable_x}Time evolution of the
    ion's wave packet along the $x$ axis, for trap parameters $U_0 =
    1.4813$ and $V_0 = 8.825$, and $\omega = 5$.  (a) Wave-packet
    width $\sigma_x$; (b) expectation value of the position
    $\braket{x}$; (c) difference between quantum and classical
    trajectories.}
\end{figure}

\begin{figure}
  \begin{center}
    \includegraphics[width=0.99\textwidth]{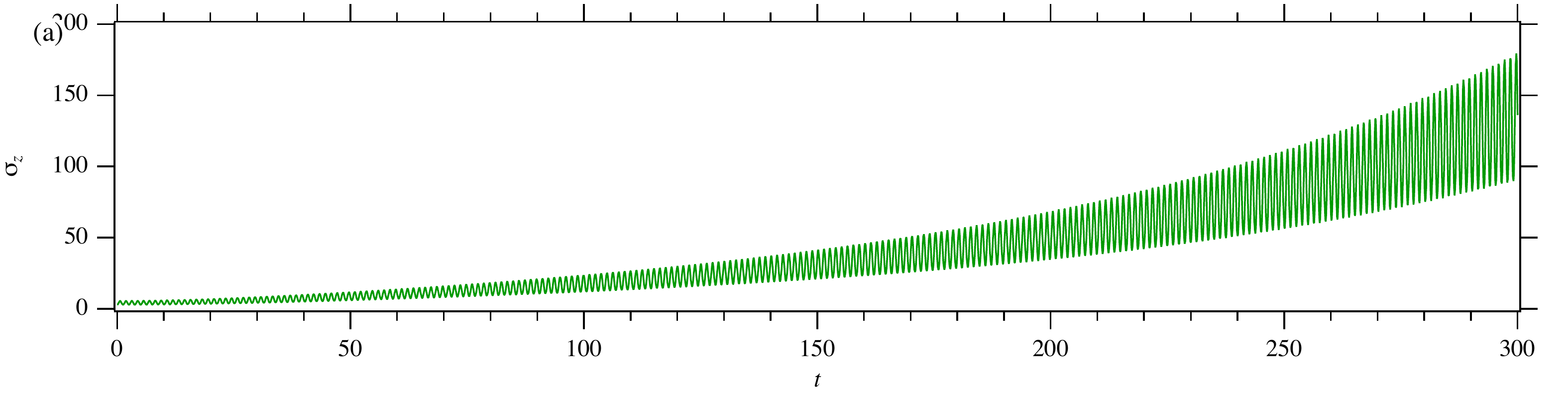}
    \includegraphics[width=\textwidth]{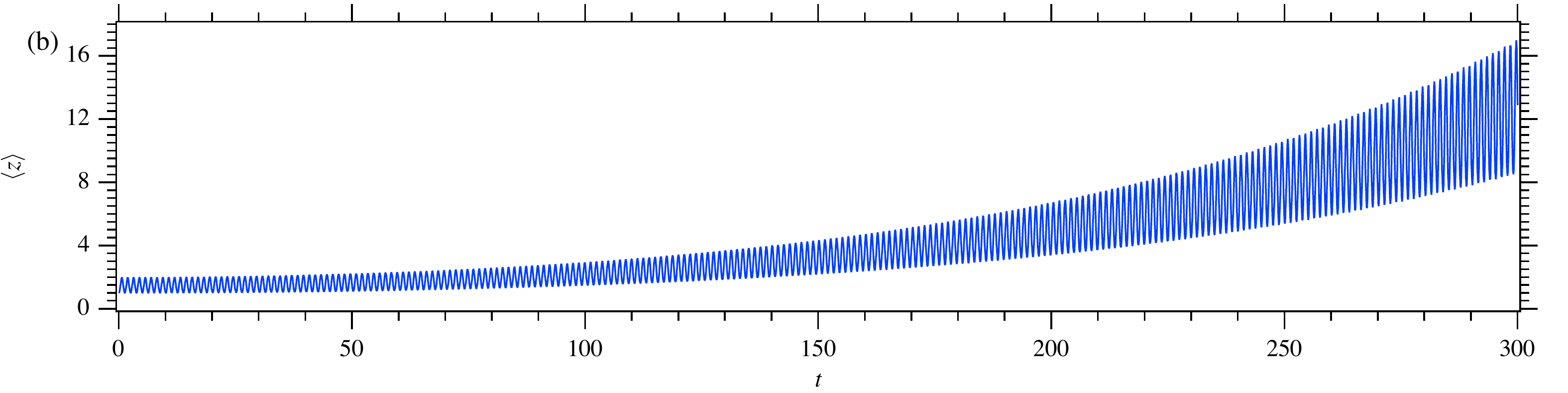}
    \includegraphics[width=\textwidth]{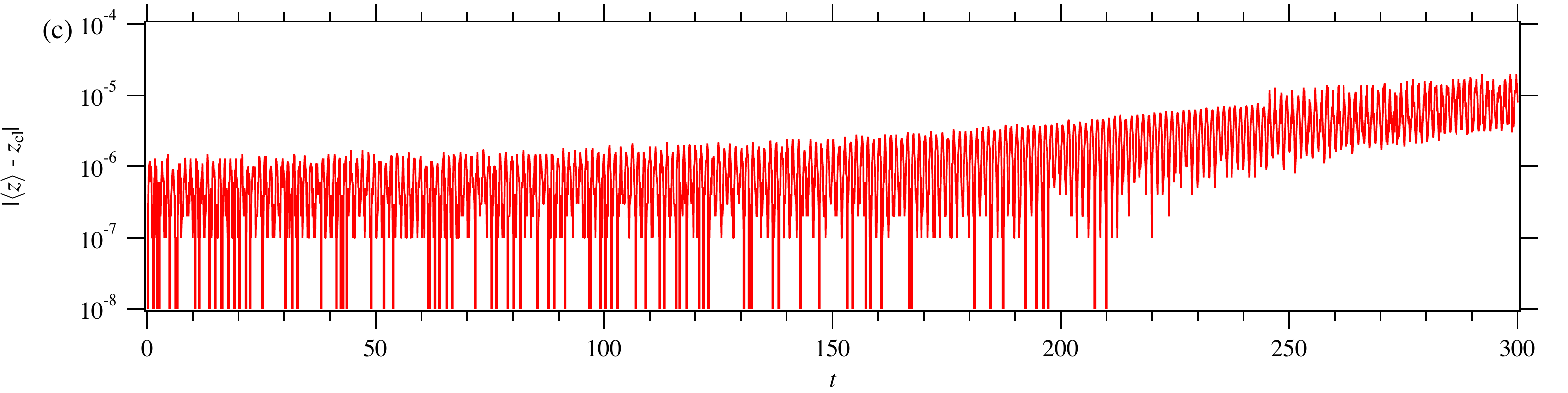}
  \end{center}
  \caption{\label{fig:unstable_z}Time evolution of the
    ion's wave packet along the $z$ axis, for trap parameters $U_0 =
    1.4813$ and $V_0 = 8.825$, and $\omega = 5$.  (a) Wave-packet
    width $\sigma_z$; (b) expectation value of the position
    $\braket{z}$; (c) difference between quantum and classical
    trajectories.}
\end{figure}

Changing slightly the value of $U_0$ to $1.4813$, the trapping should
still be stable quantum mechanically according to Figure~1 of
Reference~\cite{Wang_PRA_1995}, whereas the classical trajectory should be
unstable, see Figure~\ref{fig:stability_wang}.  The results, presented
in Figures~\ref{fig:unstable_x} and \ref{fig:unstable_z},
show clearly that, as for all our previous
results~\cite{Hashemloo_IJMPC_2016}, $\braket{x}$ and $\braket{z}$
follow exactly the classical motion, and therefore the trajectory is
\emph{not} stable.  Similarly, the width of the wave packet in both
$x$ and $z$ increases continuously, as was seen for a linear Paul trap
in Reference~\cite{Hashemloo_IJMPC_2016}.

By varying both $U_0$ and $V_0$, we have checked that, in the regions
labelled $x$-stable, $z$-unstable (or conversely, $x$-unstable,
$z$-stable) in Figure~\ref{fig:stability_wang}, the dynamical behaviour
is as expected, with stability only observed along one of the
dimensions.  In other words, as far as our simulations show, the
quantum wave-packet dynamics follow the stability criteria of the
corresponding Mathieu equation for the classical dynamics.  

We have also looked at the crossing of the boundary for quantum
stability as established by Wang \etal~\cite{Wang_PRA_1995}.  For
instance, we have compared the results along $z$ for $U_0 = 0.5$, $V_0
= 2.5$, which should be stable according to Figure~1 of
Reference~\cite{Wang_PRA_1995}, with those for $U_0 = 1$, $V_0 = 2.5$,
which lies in the $z$-unstable region.  We have found that both cases
lead to unconstrained wave packet dynamics, with no qualitative
difference between these two cases.  The presence of the boundary
presented in Figure~1 of Reference~\cite{Wang_PRA_1995} is not observable in
our simulations.

\section{Conclusion}
\label{sec:conclusion}

We have performed quantum simulations of the full dynamics of an ion
in a Paul trap, including the time-de\-pen\-dent variation of the trapping
potential.  We have shown that the centre-of-mass motion of the ion is
well reproduced by the classical equations of motion, even outside the
so-called stability region, where the trajectories are no longer
bound~\cite{Major_book_2005}.

We have revisited the work of Wang \etal~\cite{Wang_PRA_1995}, and
found that even in conditions where they have claimed that there would
be a discrepancy between quantum and classical results, we found that
the quantum trajectories were equally unstable.  This confirms that
the classical stability criterion can be applied to quantum motion,
both with respect to expectation value of the position and width of
the wave packet of the trapped ion.

We conclude that the convergence criterion considered by Wang
\etal~\cite{Wang_PRA_1995} does not inform on the stability of the
quantum trajectories, in the usual sense of the capacity of the trap
to contain the
ion~\cite{Paul_RMP_1990,Ghosh_book_1995,Major_book_2005}.
\section*{Acknowledgement(s)}

The simulations were performed on resources provided by the Swedish
National Infrastructure for Computing (SNIC) at the High Performance
Computing Center North (HPC2N).  

\section*{Disclosure statement}

No potential conflict of interest was reported by the authors.

\section*{Funding}

Financial support from Ume{\aa} University is gratefully acknowledged.

\bibliographystyle{tfo}
\bibliography{floquet,ions,methods,nist,quantum}

\clearpage


\end{document}